\documentclass[%
preprint,
 amsmath,amssymb,
 aps,
]{revtex4-1}

\usepackage{graphicx}
\usepackage{dcolumn}
\usepackage{bm}
\usepackage{amssymb,amsfonts,amsmath}
\usepackage{mathrsfs}
\usepackage{color}

\begin{document}


\title{Transition in relaxation paths in allosteric molecules:\\ Enzymatic kinetically constrained model}

\author{Tetsuhiro S. Hatakeyama}
 \email{hatakeyama@complex.c.u-tokyo.ac.jp}
\author{Kunihiko Kaneko}%
\affiliation{%
Department of Basic Science, University of Tokyo, 3-8-1 Komaba, Meguro-ku, Tokyo 153-8902, Japan
}%

\date{\today}

\begin{abstract}
A hierarchy of timescales is ubiquitous in biological systems, where enzymatic reactions play an important role because they can hasten the relaxation to equilibrium.
We introduced a statistical physics model of interacting spins that also incorporates enzymatic reactions to extend the classic model for allosteric regulation.
Through Monte Carlo simulations, we found that the relaxation dynamics are much slower than the elementary reactions and are logarithmic in time with several plateaus, as is commonly observed for glasses.
This is because of the kinetic constraints from the cooperativity via the competition for an enzyme, which has different affinity for molecules with different structures.
Our model showed symmetry breaking in the relaxation trajectories that led to inherently kinetic transitions without any correspondence to the equilibrium state.
In this paper, we discuss the relevance of these results for diverse responses in biology.
\end{abstract}

\pacs{Valid PACS appear here}
\maketitle


\section{Introduction}

Biological systems are known to have a hierarchy of timescales \cite{Phillips2008}.
Ordinarily, the timescale of biochemical reactions is of the subsecond order, that of organisms' behaviors is of the order of seconds to hours, and that of lifespans is of the order of years.
How organisms fill the gaps between such timescales remains one of the most important problems in biophysics.

As long as the biochemical system follows the Michaelis--Menten kinetics and the concentration of a substrate is saturated, which is ordinary in cells, the gap of timescales between biochemical reactions and organisms' behaviors is hardly filled.
Recent studies, however, reported that the kinetics of the multisite modification of proteins does not always follow ordinary Michaelis--Menten kinetics \cite{Mylona2016, Salazar2009, Zhou2015, Hatakeyama2012, Hatakeyama2014}.
For example, in the Erk/Elk-1 signaling pathway, the timescales of the phosphorylation reactions are broadly distributed among multiple sites, and the phosphorylation speed of each site depends on both the site itself and the modification of other sites \cite{Mylona2016}.
Extensive theoretical studies have shown that multisite modification and the competition for limited enzyme abundances can change the kinetics as well as the steady-state modification level \cite{Salazar2009, Hatakeyama2012, Hatakeyama2014}.
Notably, sequential multisite modification has been reported to generate a variety of timescales, some of which are much slower than the enzymatic turnover rate \cite{Hatakeyama2014}.

Such regulation of the modification kinetics, including the slow dynamics, is considered to result from intermolecular cooperativity.
Although a dynamical-system model with chemical kinetics has previously been proposed for the average relaxation process, a model and analysis that go beyond dynamical systems are required to reveal the intermolecular cooperativity and the fluctuations in the slow relaxation process.

Concepts from statistical physics may be useful for investigating the slow biochemical dynamics and its fluctuation.
Such slow dynamics have been extensively and intensively studied with regard to the physics of glasses \cite{Debenedetti2001}.
In kinetically constrained models (KCMs), relaxation to the equilibrium is kinetically suppressed without thermodynamic metastable states exist in the energy landscape \cite{Ritort2003}.
A promising mechanism exists for kinetically slowed-down processes without resorting to the existence of multiple metastable states, namely, controlling the enzyme abundance. Because the reaction rate is controlled enzymatically, the lack of an enzyme may suppress the corresponding reactions.
Despite the possible relevance of the kinetic constraint concept to biochemical processes, it has not been fully explored due to the lack of a KCM for biological systems.

\begin{figure}[tbhp]
\centering
\includegraphics[]{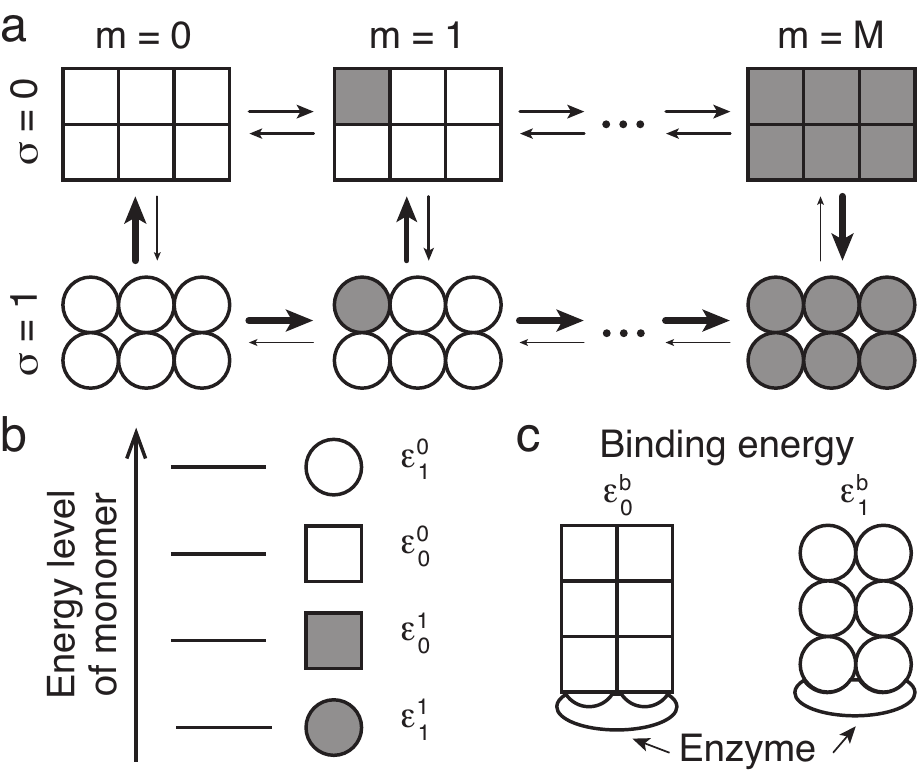}
\caption{
Schematics of the enzymatic MWC model.
(a) All states that a single MWC molecule can take.
$\sigma$ represents the T or R state and $m$, the number of modifications.
(b) Energy level of each monomer.
(c) Binding energies between the enzyme and the MWC molecules with different states.
The binding energy depends on a molecular structure, $\sigma$.
\label{fig:model}
}
\end{figure}

In this study, to uncover a relationship between the slow dynamics in biology and the kinetic constraint, we introduce an enzymatic kinetically constrained model (eKCM) by adopting the Monod--Wyman--Changeux (MWC) model for multiple modifications of the protein state (Fig.~\ref{fig:model}), which is the classic established model for concerted allosteric regulation \cite{Monod1965}.
The essence of allostery is represented by a coupling between modification and structure of a molecule.
Here, we consider two types allosteric effect: (i) {\it energetic effect} and (ii) {\it enzymatic (kinetic) effect}, the former has been considered in statistical-physics models \cite{Marzen2013} and the latter had been never considered and is required for eKCM.
We demonstrate that even if both effects accelerate reactions, the eKCM counterintuitively showed a slow relaxation to the equilibrium state.
A typical time course for relaxation showed multiple plateaus, where the modification progress was transiently frozen far from equilibrium.

\section{Model}

The model includes both modification of multiple monomers and a large structural change between the tense (T) and relaxed (R) states.
We here represent each modification and each structural state as two types of ``spins''; that is, each molecule has $M$ modification spins and a single structural spin taking a down or up state, where $M$ is the number of monomers and set as 6 unless otherwise noted.
Thus, each molecule has $2^{M+1}$ states.
The modification spin $s_i$ flips from 0 to 1 when each site is modified, where $i$ is $1 \leq i \leq M$, and the total number of up spins is denoted as $m$.
The structural spin $\sigma$ flips between the T ($\sigma = 0$) and R ($\sigma = 1$) state with a structural change.
The internal energy of a single molecule is defined as the summation of energies of all modification spins.
Thus, the Hamiltonian of the single molecule is given as
\begin{equation}
\mathscr{H} (\sigma, \{s_i\}) = \left[ m \epsilon^1_{\sigma} (h) + (M - m) \epsilon^0_{\sigma} \right] \label{eq:Hamiltonian},
\end{equation}
where $\epsilon^1_{\sigma} (h)$ and $\epsilon^0_{\sigma}$ are the energies of the up and down spins, respectively, of the $\sigma$ state molecule.
$\epsilon^1_{\sigma} (h)$ is a function of $h$, which is the ``chemical field'' derived from the concentration of the coenzyme required for the transfer of functional groups, and it is set as constant without losing generality.
$\{s_i\}$ is a set of modification spins.
The number of molecules in the present system is fixed at $N$.
We set $N$ as 100 unless otherwise noted.
Therefore, the partition function is given as
\begin{equation}
Z = \left[ \sum_{\sigma = 0, 1} \sum_{i} \binom{M}{i} \exp \left\{-\beta ((M - i) \epsilon_0^{\sigma} + i \epsilon_1^{\sigma}) \right\} \right]^N.
\end{equation}
We introduce two quantities: the fraction of unmodified monomers $\mathcal{U}$ and the T-state molecule ratio $\mathcal{T}$. These are defined as $1 - \Sigma_j m_j / (N M)$ and $N_0 / N = 1 - N_1 / N$, respectively, where $m_j$ is the modification level $m$ of $j$th molecule, $N_0$ and $N_1$ are the numbers of T- and R-state molecules, respectively.
Such quantities in the equilibrium state are easily derived from the partition function because there is no interaction term among the molecules in the Hamiltonian.
The two types of allosteric effects to accelerate the reaction are formulated as follows.

\noindent
(i) {\it Energetic effect}

If a molecule is modified at many modification sites, such a molecule tends to change from the T to R state. The modification sites of R-state proteins are easier to modify than those of T-state proteins.
Therefore, the energy of each modification spin has to satisfy the inequality $\epsilon^0_1 > \epsilon^0_0 > \epsilon^1_0 > \epsilon^1_1$ (Fig. \ref{fig:model}(b)). Here, $\epsilon^0_1$, $\epsilon^0_0$, $\epsilon^1_0$, and $\epsilon^1_1$ are set to 4, 3, 2, 1, respectively.

By considering the detailed balance condition, the transition probability of each protein state is given as
\begin{align}
&\frac{p(\sigma, \{1, \cdots, s_M \} | \sigma, \{0, \cdots, s_M \})}{p(\sigma, \{0, \cdots, s_M \} | \sigma, \{1, \cdots, s_M \})} = \frac{\exp(- \beta \epsilon^1_{\sigma})}{\exp(- \beta \epsilon^0_{\sigma})}, \label{eq:DB_modi} \\
&\frac{p(1, \{s_i\} | 0, \{s_i\})}{p(0, \{s_i\} | 1, \{s_i\})} = \frac{\exp[- \beta \{m \epsilon^1_1 + (M - m) \epsilon^0_1\} ]}{\exp[- \beta \{m \epsilon^1_0 + (M - m) \epsilon^0_0\} ]}, \label{eq:DB_state}
\end{align}
where $p(\sigma, \{1, \cdots, s_M \} | \sigma, \{0, \cdots, s_M \})$ and $p(\sigma, \{0, \cdots, s_M \} | \sigma, \{1, \cdots, s_M \})$ are the transition probabilities for the modification and non-modification of the 1st modification spin of a $\sigma$-state molecule, respectively.
The same transition probabilities are adopted for the other modification spins.
$p(1, \{s_i\} | 0, \{s_i\})$ and $p(0, \{s_i\} | 1, \{s_i\})$ are the transition probabilities for the structural change from $\sigma = 0$ to $\sigma = 1$ and $\sigma = 1$ to $\sigma = 0$, respectively, when modification state is $\{s_i\}$.

When $m$ is small, the structure tends to be $\sigma = 0$ because $\epsilon^0_1 > \epsilon^0_0$, whereas the structure tends to be $\sigma = 1$ because $\epsilon^1_0 > \epsilon^1_1$ for large $m$.
The transition probabilities from $\sigma = 0$ to 1 and $\sigma = 1$ to 0 are identical for $m = M/2$.
Here, we assume that the structural change can always occur within the characteristic time of that when the microscopic energy decreases, i.e., we adopt the Metropolis method for structural change.

For the modification, the activation energy is set to be equal to $\epsilon^0_1$ for all modification reactions for simplicity.
Hence, the modification of the R-state molecule has no energy barrier, whereas that of the T-state molecule has an energy barrier.

\noindent
(ii) {\it Enzymatic (kinetic) effect}

Although the enzyme works as a catalyst for modification and does not change the detailed balance condition, competition for the enzyme among molecules introduces a kinetic effect.
By assuming that the timescale of enzyme binding is much faster than that of modification and state change, the binding reaction can be eliminated adiabatically.
Then, the kinetics of the modification are governed by the product of the binding probability ($P^{\rm b}_{\sigma}$) and the activation probability to go across the energy barrier.
Hence, the enzymatic effect is represented by the changes in $P^{\rm b}_0$ and $P^{\rm b}_1$ following the structural change.

To accelerate the reaction, the R-state molecule tends to bind the enzyme more and is modified faster than the T state, i.e., $P^{\rm b}_1$ is larger than$P^{\rm b}_0$.
Thus, the T- and R-state molecules have different binding energies with the enzyme of $\epsilon_0^{\rm b}$ and $\epsilon_1^{\rm b}$, respectively, where $\epsilon_0^{\rm b}$ is lower than $\epsilon_1^{\rm b}$ ($\epsilon_0^{\rm b} = 0$ and $\epsilon_1^{\rm b} = 10$).
Under the assumption that only a single enzyme can bind to the molecule, the transition probability for the modification under the detailed balance condition (Eq.~(\ref{eq:DB_modi})) is 
\begin{align}
&p(\sigma, \{1, \cdots, s_M \} | \sigma, \{0,  \cdots, s_M \}) = P^{\rm b}_{\sigma} \exp(-\beta \{\epsilon^0_1 - \epsilon^0_{\sigma} \}), \nonumber \\
&p(\sigma, \{0, \cdots, s_M \} | \sigma, \{1, \cdots, s_M \}) = P^{\rm b}_{\sigma} \exp(-\beta \{\epsilon^0_1 - \epsilon^1_{\sigma} \}), \nonumber \\
&P^{\rm b}_{\sigma} = \frac{<n_{\sigma}>}{N_{\sigma}} = \frac{\exp(\beta \mu)}{\exp(-\beta \epsilon^{\rm b}_{\sigma}) + \exp(\beta \mu)},
\end{align}
where $n_{\sigma}$ is the number of enzymes that bind to the $\sigma$-state molecule (see the Supplemental Material for the derivation) .
Fig. S1 shows the binding probabilities calculated thus far.

We set the timescales of the modification flip and state flip as $\tau_{\rm m}$ and $\tau_{\rm s}$, respectively, where $\tau_{\rm m}$ is longer than $\tau_{\rm s}$ ($\tau_{\rm s} = 1.0$, $\tau_{\rm m} = 10.0$).
We set the initial condition such that all molecules are in the $\sigma = 0$ and $m = 0$ states, and we investigated the relaxation dynamics to the
equilibrium state using the Monte Carlo method.

\section{Results}

\begin{figure}[tbhp]
\centering
\includegraphics[]{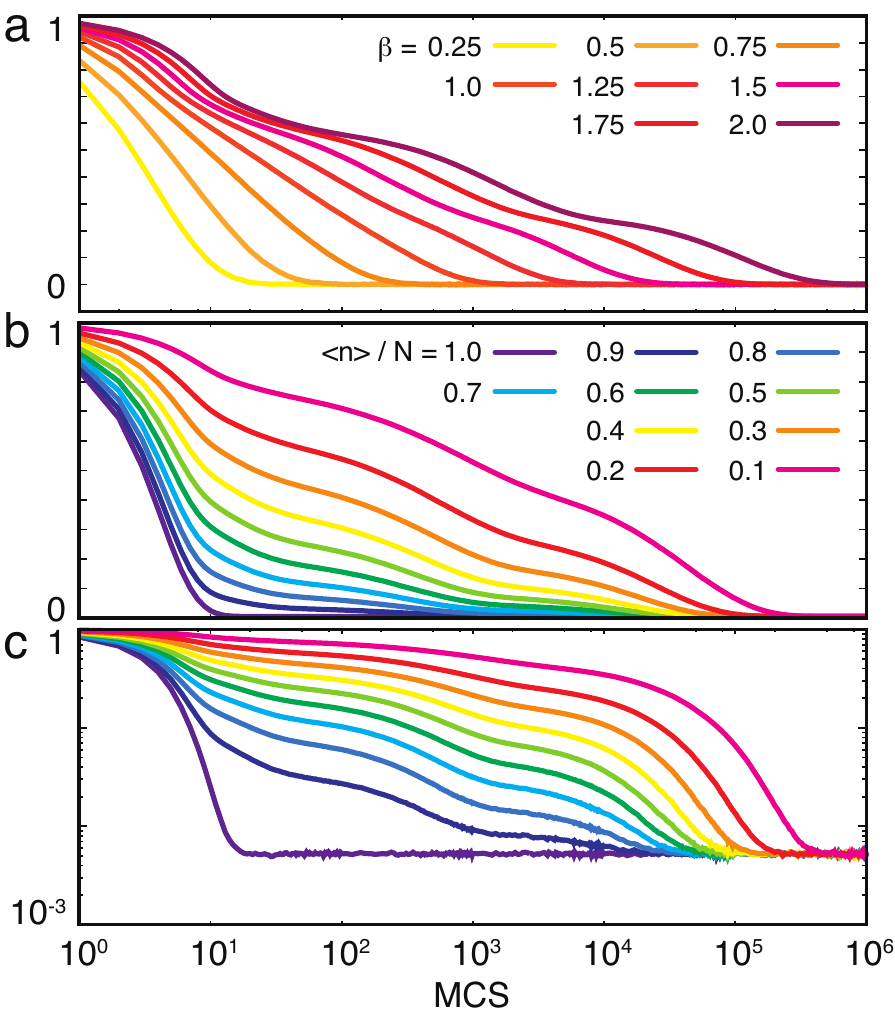}
\caption{
Time evolution of the average modification in the eMWC against the logarithmic Monte Carlo step.
(a) Relaxation of the average modification to the equilibrium at various temperatures.
Since $\mathcal{U}_{\rm eq}$ depends on the temperature, the normalized ratio $(<\mathcal{U}> _{\rm ens}- \mathcal{U}_{\rm eq})/(1 - \mathcal{U}_{\rm eq})$ was plotted by setting $<n>/N$ at 0.2.
Different color lines indicate time courses with different values of $\beta$.
(b) Relaxation of the average modification to the equilibrium for various values of $<n>$.
The time course of $<\mathcal{U}>_{\rm ens}$ was plotted by setting $\beta$ at 1.75.
The different line colors correspond to different values of $<n>/N$.
($\mathcal{U}_{\rm eq}$ is independent of $<n>$).
Each line is an ensemble average of 1000 samples.
(c) Logarithm of $<\mathcal{U}>_{\rm ens}$ plotted for (b).
\label{fig:spin_time_course}
}
\end{figure}

First, we calculated the relaxation dynamics of $\mathcal{U}$ and $\mathcal{T}$.
Although $<\mathcal{U}>_{\rm ens}$ and $<\mathcal{T}>_{\rm ens}$, where $<>_{\rm ens}$ is the ensemble average, finally relaxed to the equilibrium values $\mathcal{U}_{\rm eq}$ and $\mathcal{T}_{\rm eq}$, respectively, their time courses varied depending on the temperature $1 / \beta$ and $<n>$, which is the average number of enzymes binding to a substrate (Fig.~\ref{fig:spin_time_course}).
When the temperature was high, $<\mathcal{U}>_{\rm ens}$ decreased exponentially with time with no plateau.
As the temperature decreased, the relaxation slowed down and decreased logarithmically with time.
Two plateaus appeared as the temperature decreased further (see Fig.~\ref{fig:spin_time_course}(a)).
The two plateaus were clearly discernible when $<n>$ was reduced below $<n>/N = 0.95$.
The relaxation of $<\mathcal{T}>_{\rm ens}$ also showed a similar dependence on $<n>$ and the temperature, as shown in Fig. S2.
Such slow dynamics with plateaus have often been observed in glasses \cite{Debenedetti2001}.

\begin{figure}[tbhp]
\centering
\includegraphics[]{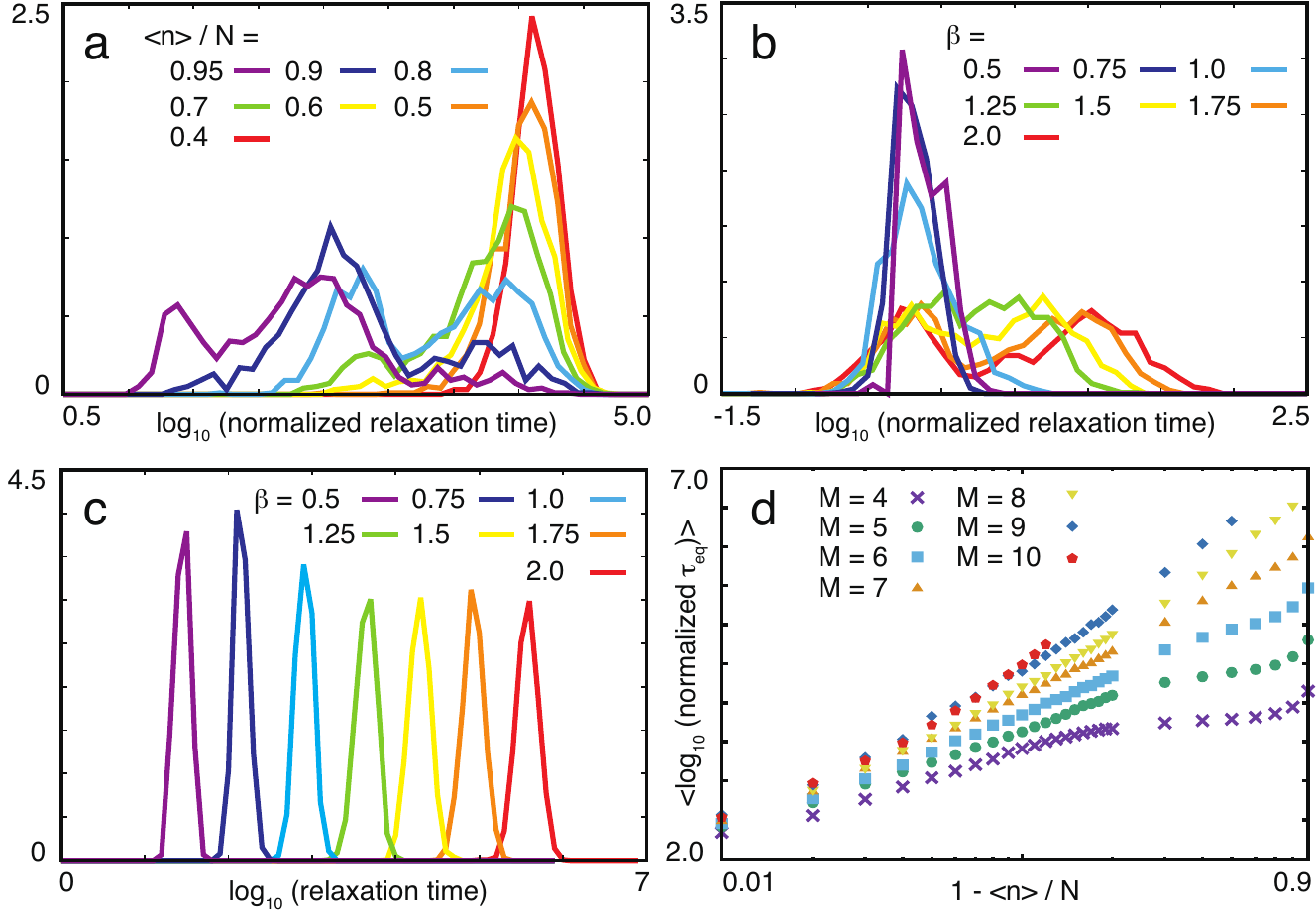}
\caption{
Distributions of the relaxation time and the variance of the logarithmic relaxation time.
Probability distribution of the logarithmic normalized relaxation time at various $<n>$ values with fixed temperature at $\beta = 1.75$ (a), and at various temperatures with $<n>/N$ fixed at 0.8 (b) and 0.2 (c), respectively.
The logarithmic relaxation time was calculated as the base-10 logarithm of $\tau_{\rm eq}$.
Plots are rescaled by $\tau_{\rm eq} <n>^{-1}$ for (a) and rescaled by $\tau_{\rm eq} e^{-4 \beta}$ for (b).
Different color lines indicate the probability distributions under different parameters.
(d) Averaged normalized relaxation time for different number of modification sites $M$ indicated by different symbols.
The plot is rescaled in the similar way as in (a).
\label{fig:relax_time_dist}
}
\end{figure}

To reveal the mechanism of the anomalous parameter dependence, we analyzed the relaxation-time distribution over the samples.
In the region where the relaxation time showed anomalous parameter dependence, the relaxation-time distribution changed from unimodal to multimodal (Fig.~\ref{fig:relax_time_dist}).
The multiple peaks that emerged are named the first, second, and third peaks in ascending order of relaxation time.

When $<n>$ was varied at a fixed temperature, the positions of the peaks changed in proportion to $<n>^{-1}$ (see Fig.~\ref{fig:relax_time_dist}(a)).
Thus, we studied the distribution of the relaxation time normalized by $<n>^{-1}$.
When $<n>/N$ was close to 1, two peaks were observed.
As $<n>$ decreased, the first peak disappeared and was replaced by the third one.
Finally, the second peak disappeared completely at $<n>/N = 0.4$.
This change in the distribution is similar to the first-order phase transition in equilibrium thermodynamics.
Indeed, as $M$ was increased, the divergence of the relaxation-time against $<n>/N \rightarrow 1$ was steeper (Fig.~\ref{fig:relax_time_dist}(d)) and its variance increased (Fig. S3).
This suggests that in the limit of $M \rightarrow \infty$ and $N \rightarrow \infty$, the change in the relaxation time is similar to the phase transition in the context of equilibrium thermodynamics.
Actually, as $N$ increased, the divergence of the relaxation time was steeper even in the case of $N = 6$ (see Fig. S4).
It should be recalled that the same equilibrium state was reached over all samples independent of the relaxation courses, and the transition here is with regard to the relaxation trajectories rather than the quantities in thermodynamic equilibrium.

The temperature dependence of the relaxation-time distribution when $<n>/N$ was fixed close to 1 differed from its $<n>$ dependence (see Fig.~\ref{fig:relax_time_dist}(b)).
In this case, the position of the first peak changed in proportion to $\exp(4 \beta)$, as explained later.
At high temperatures, the relaxation-time distribution was unimodal.
This distribution broadened with decreasing temperature at around $\beta = 1.0$, where it became bimodal, and the distance between these two peaks increased with decreasing temperature.
This change in the relaxation-time distribution is similar to the second-order phase transition in equilibrium, whereas when $<n>/N$ is small, the relaxation-time distribution shows not a transition but a crossover (Fig.~\ref{fig:relax_time_dist}(c)).
Indeed, the variance of the logarithmic relaxation time is large at around $<n>/N = 1$ (Fig. S5(a)).

\begin{figure}[tbhp]
\centering
\includegraphics[]{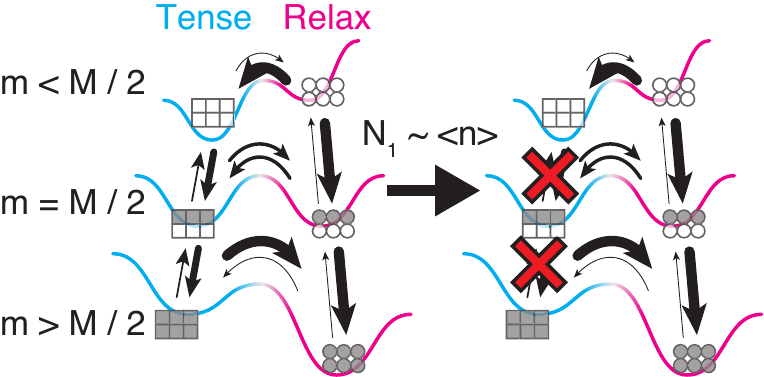}
\caption{
Schematic showing the energy landscape of the eKCM.
(Left)
If the number of R-state molecules is lower than that of the enzyme, the modification reaction of T-state molecules (up and down arrows) can occur and progress until half of the modification sites are modified.
Then, when more than half of the sites are modified, such molecules tend to be in the R state (lower right arrow).
(Right)
When the number of R-state molecules exceeds that of the enzyme through modification, the R-state molecules monopolize the enzyme, and the modification reaction of T-state molecules is kinetically inhibited (red crosses).
Therefore, the structural changes from the T to R state in the less modified molecules (upper right arrow) are rate-limiting.
\label{fig:energy_scheme}
}
\end{figure}

This transition is caused by kinetic constraint from the competition for the enzyme among R- and T-state molecules.
As the relaxation progresses and the number of R-state molecules reaches almost the same level as that of the enzyme, the R-state molecules monopolize the enzyme due to the positive allosteric effect (see Fig. S1).
Then, further progress in the modification reactions of the T-state molecules is suppressed (see Fig.~\ref{fig:energy_scheme}).
Then, the transition from the T- to the R-state is rate-limiting with three possible steps, $m = 0$, 1, and 2, having the energy barriers 6, 4, and 2, respectively.
Thus, the temperature dependence of the transition rate follows $\exp (6 \beta)$ ($m=0$), $\exp (4 \beta)$ ($m=1$), and $\exp (2 \beta)$ ($m=2$).

When the number of molecules is finite, the rate-limiting step depends on the distribution of the phosphorylation level of the T-state molecules at the start of the kinetically constrained condition.
As long as the $(m = 0, \sigma = 0)$ molecule exists, the transition from $(m = 0, \sigma = 0)$ to $(m = 0, \sigma = 1)$ is rate-limiting, whereas if it is not included, its transition from $(m = 1, \sigma = 0)$ to $(m = 1, \sigma = 1)$ is rate-limiting as long as the $m=1$ molecule exists, and so forth.
These three rate-limiting steps, therefore, correspond to the third, second, and first peaks in Fig.~\ref{fig:relax_time_dist}(a), respectively, and the three plateaus of relaxation (Fig.~\ref{fig:spin_time_course}).
Indeed, only the samples in the peak with the longest relaxation time and having three plateaus had molecules in the $m = 0$ state (see Fig. S6).
This suggests that the fluctuation in the distribution of molecular states at the beginning of relaxation affects the entire process in the trajectory.

\begin{figure}[tbhp]
\centering
\includegraphics[]{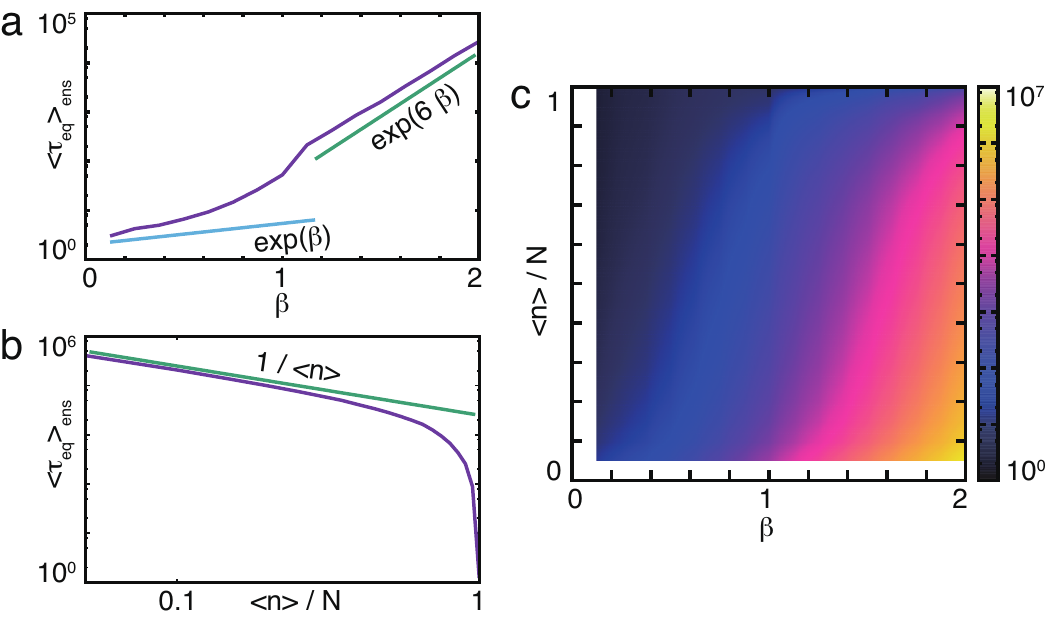}
\caption{
Relaxation times at various temperatures and $<n>$ values.
(a) Dependence of the relaxation time on the temperature.
The relaxation time is defined as the ensemble average of $\tau_{\rm eq}$ when $\mathcal{T}$ falls below the analytically calculated value $\mathcal{T}_{\rm eq}$ because $\mathcal{T}$ approaches equilibrium from above.
The blue and green lines indicate lines proportional to $\exp(\beta)$ and $\exp (6 \beta)$, respectively.
$<n>/N$ was set at 0.8.
(b) Dependence of the relaxation time on $<n>/N$.
The green line indicates a line proportional to $<n>^{-1}$.
$\beta$ was set at 1.75.
(c) Logarithmic average of the relaxation time.
$\log_{10} <\tau_{\rm eq}>_{\rm ens}$ against $\beta$ and $<n>/N$ is shown using the colors in the side bar.
\label{fig:relax_time}
}
\end{figure}

Reflecting such transition, the average relaxation time shows anomalous dependence on $1/\beta$ (Fig.~\ref{fig:relax_time}).
When $<n>$ was fixed and the temperature was varied, the relaxation time did not follow the standard Arrhenius form (see Fig.~\ref{fig:relax_time}(b)), similar to the glass.
Because the energy barrier for the modification of the T-state molecule was set to unity while the modification of the R-state molecule was temperature-independent, the relaxation time would normally be expected to be proportional to $\exp(\beta)$.
Indeed, at high temperatures, the relaxation time approximately followed $\exp(\beta)$.
However, as the temperature decreased, the rate of increase in the relaxation time against the temperature was enhanced and reached its maximum at around $\beta = 1$.
The relaxation time showed a bending point around $\beta \sim 1$, and for low temperatures, it approximately obeyed $\exp(6 \beta)$.
It suggests that the transition from the $(m = 0, \sigma = 0)$ to $(m = 0, \sigma = 1)$ is rate-limiting, whose activation energy is six times the energy barrier for each modification (see Fig.~\ref{fig:model}).

The relaxation time also shows anomalous dependence on $<n>$ around $<n>/N = 1$ (see Fig.~\ref{fig:relax_time}(b)), where the transition of distribution of relaxation time is observed.
In the ordinal chemical reaction, the relaxation time changes in proportion with the number of the complex $<n>$ in accordance with the binding probability.
Indeed, for lower $<n>$, the relaxation time was proportional to the inverse of $<n>$.
However, it was further prolonged beyond $<n>^{-1}$ as $<n>/N$ decreased to approach 1.

\section{Discussion}

Here, we propose a statistical physics model to adopt the kinetic constraint concept to biochemical systems.
In our model, the kinetic constraint is autonomously imposed by competition for the enzyme among T- and R-state molecules, resulting in the glassy relaxation.
In consistency with the standard KCM, the present kinetic constraint is generated and controlled by the enzyme abundance, which works as a parameter for the phase transition; therefore, it is termed as eKCM.

The relaxation of modification of the T-state molecules is frozen by the competition for the enzyme, whereas the R-state molecules are partially equilibrated.
This is similar to the dynamical heterogeneity, which is important in glass theory \cite{Berthier2011}.
The number of frozen molecules depends on the history and differs for each replicate (Figs. S7 and S8).
The variance in modification level approaches $M$ through the relaxation (see Fig. S7).
This reflects the MWC-type allostery, where all the $M$ monomers flip their molecular structure cooperatively.

The eKCM exhibited the transition in the relaxation paths to equilibrium, depending on the temperature and the enzyme concentration.
Following the transition of the trajectories, slow relaxation and plateaus appeared, as in the glass transition.
Most of the KCMs studied thus far, however, do not show the transition of paths with change in temperature \cite{Ritort2003}, and needless to say, there is no external parameter corresponding to the enzyme abundance.
In the eKCM, the strength of the cooperativity in kinetics depends on the enzyme abundance and temperature.
When the amount of enzyme is greater than that of the substrate, there is no competition, and glassy behavior does not appear.
The competition for a limited amount of the enzyme introduces interactions among molecules, resulting in the transition to a state with heterogeneity in the relaxation paths.

Our study also demonstrated that microscopic fluctuation in molecular states can be amplified to a large variation in the relaxation time.
Although the noise in chemical concentrations has recently attracted much attention from many physicists and biologists \cite{Elowitz2002}, there has been little study on the fluctuations in the relaxation paths in biology.
They can be easily observed experimentally by using a biochemical reaction in a liposome or emulsion, as larger fluctuations are expected for a system with a small number of molecules.

In recent decades, the slow relaxation process has also attracted much interest among biophysicists.
In bacterial chemotaxis, chemoreceptors, which are often described by an MWC-type model \cite{Asakura1984, Lan2016}, are known to form clusters with each other and show a slow logarithmic change in their structure with time in response to the addition and removal of a ligand \cite{Frank2013}.
Other example is the phosphorylation of PER2 by a kinase, CKI$\epsilon$/$\delta$, which determines the period of the mammalian circadian clock \cite{Isojima2009}.
Interestingly, CKI$\epsilon$/$\delta$ tends to rebind its own catalytic products, and mutant mice with lower rebinding activity showed a shorter period of circadian rhythm \cite{Shinohara2017}, which corresponds to our results.
We expect that the experiments with above systems will demonstrate the slow relaxation of modifications as well as large variance of relaxation time, and deepen our understanding of the relationship between the regulation of biological timescales and glass theory in physics.

\begin{acknowledgments}
We thank Atsushi Ikeda for critical reading of the manuscript and Yasushi Okada and Tom Shimizu for helpful discussions.
This work was partially supported by Grants-in-Aid for Scientific Research, KAKENHI, grant number 17H05758 and 15H05746, and Grant-in-Aid for Scientific Research on Innovative Areas from the Ministry of Education, Culture, Sports, Science and Technology (MEXT) of Japan, grant number 17H06386.
\end{acknowledgments}


\begin{thebibliography}{99}
\bibitem{Phillips2008} R. Phillips, J. Kondev, and J. Theriot, {\it Physical Biology of the Cell} (Garland Science, Taylor \& Francis Group, New York, 2008).
\bibitem{Mylona2016} A. Mylona, F. X. Theillet, C. Foster, T. M. Cheng, F. Miralles, P. A. Bates,  P. Selenko, and R. Treisman, Opposing effects of Elk-1 multisite phosphorylation shape its response to ERK activation, {\it Science} \textbf{354}, 6309 (2016).
\bibitem{Salazar2009} C. Salazar and T. H\"{o}fer, Multisite protein phosphorylation -- From molecular mechanisms to kinetic models, {\it FEBS J.} \textbf{276}, 12 (2009).
\bibitem{Hatakeyama2012} T. S. Hatakeyama and K. Kaneko, Generic temperature compensation of biological clocks by autonomous regulation of catalyst concentration, {\it Proc. Natl. Acad. Sci. USA} \textbf{109}, 21 (2012).
\bibitem{Hatakeyama2014} T. S. Hatakeyama and K. Kaneko, Kinetic memory based on the enzyme-limited competition, {\it PLoS Comput. Biol.} \textbf{10}, 8 (2014).
\bibitem{Zhou2015} M. Zhou, J. K. Kim, G. W. L. Eng, D. B. Forger, and D. M. Virshup, A period2 phosphoswitch regulates and temperature compensates circadian period, {\it Mol. Cell} \textbf{60}, 1 (2015).
\bibitem{Debenedetti2001} P. G. Debenedetti and F. H. Stillinger, Supercooled liquids and the glass transition, {\it Nature} \textbf{410}, 6825 (2001).
\bibitem{Ritort2003} F. Ritort and P. Sollich, Glassy dynamics of kinetically constrained models, {\it Adv. Phys.} \textbf{52}, 4 (2003).
\bibitem{Monod1965} J. Monod, J. Wyman, and J. -P. Changeux, On the nature of allosteric transitions: A plausible model, {\it J. Mol. Biol.} \textbf{12} (1965).
\bibitem{Marzen2013} S. Marzen, H. G. Garcia, and R. Phillips, Statistical mechanics of Monod--Wyman--Changeux (MWC) models, {\it J. Mol. Biol.} \textbf{425}, 9 (2013).
\bibitem{Berthier2011} L. Berthier, G. Biroli, J. -P. Bouchaud, L. Cipelletti, and W. van Saarloos (Eds.), {\it Dynamical Heterogeneities in Glasses, Colloids, and Granular Media} (Oxford University Press, New York, 2011).
\bibitem{Elowitz2002} M. B. Elowitz, A. J. Levine, E. D. Siggia, and P. S. Swain, Stochastic gene expression in a single cell, {\it Science} \textbf{297}, 5584 (2002).
\bibitem{Asakura1984} S. Asakura and H. Honda, Two-state model for bacterial chemoreceptor proteins: The role of multiple methylation, {\it J. Mol. Biol.} \textbf{176}, 3 (1984).
\bibitem{Lan2016} G. Lan and Y. Tu, Information processing in bacteria: Memory, computation, and statistical physics: A key issues review, {\it Reports. Prog. Phys.} \textbf{79}, 5 (2016).
\bibitem{Frank2013} V. Frank and A. Vaknin, Prolonged stimuli alter the bacterial chemosensory clusters, {\it Mol. Microbiol.} \textbf{88}, 3 (2013).
\bibitem{Isojima2009} Y. Isojima et al., CKI$\epsilon$/$\delta$-dependent phosphorylation is a temperature-insensitive, period-determining process in the mammalian circadian clock, {\it Proc. Natl. Acad. Sci. USA} \textbf{106}, 37 (2009).
\bibitem{Shinohara2017} Y. Shinohara et al., Temperature-sensitive substrate and product binding underlie temperature-compensated phosphorylation in the clock, {\it Mol. Cell} \textbf{67}, 5 (2017).
\end{thebibliography}
\end{document}


\maketitle

\section{Derivation of the binding probability of the enzyme}

We assume that only a single enzyme can bind to the molecule.
The T- and R-state molecules have different binding energies with the enzyme of $\epsilon_0^{\rm b}$ and $\epsilon_1^{\rm b}$, respectively, where $\epsilon_0^{\rm b}$ is lower than $\epsilon_1^{\rm b}$ for the positive allosteric effect.
Therefore, the grand partition function and the average number of binding enzymes $<n>$ in equilibrium are represented in a similar manner as Langmuir's adsorption isotherm \citep{Phillips2008} and are given as
\begin{eqnarray}
\Xi = \left\{1 + \exp(\beta (\epsilon^{\rm b}_0 + \mu)) \right\}^{N_0} \left\{1 + \exp(\beta (\epsilon^{\rm b}_1 + \mu)) \right\}^{N_1}, \\
<n> = N_0 \frac{\exp(\beta \mu)}{\exp(-\beta \epsilon^{\rm b}_0) + \exp(\beta \mu)} + N_1 \frac{\exp(\beta \mu)}{\exp(-\beta \epsilon^{\rm b}_1) + \exp(\beta \mu)}. \label{eq:avg_n}
\end{eqnarray}

For both {\it in vivo} and {\it in vitro} reactions, $N$ and $n$ are almost constant and are the control parameters of the system.
Here, we assume that $<n>$ is the same as $n$, which is the total number of enzymes in the system, and that the chemical potential $\mu$ is a function of $<n>$, $N_0$, and $N_1 = N - N_0$.
This assumption is justified when $N$ and $n$ are sufficiently large, i.e., at the thermodynamic limit.
Then, $\exp(\beta \mu)$ is given as
\begin{equation}
\exp(\beta \mu) = \frac{-A + \sqrt{A^2 - 4 <n> (<n> - N) \exp(-\beta (\epsilon^{\rm b}_0 + \epsilon^{\rm b}_1))}}{2 (<n> - N)}, \label{eq:chemical_potential}
\end{equation}
where $A = (<n> - N + N_0) \exp(-\beta \epsilon^{\rm b}_0) + (<n> - N_1) \exp(-\beta \epsilon^{\rm b}_1)$.
In the limit $<n> \rightarrow N$, $\mu$ approaches $-\infty$, and all molecules bind to the enzyme.
The arguments so far can be summarized to give the binding probability
\begin{equation}
P^{\rm b}_{\sigma} = \frac{<n_{\sigma}>}{N_{\sigma}} = \frac{\exp(\beta \mu)}{\exp(-\beta \epsilon^{\rm b}_{\sigma}) + \exp(\beta \mu)},
\end{equation}
where $n_{\sigma}$ is the average number of enzymes that bind to the $\sigma$-state molecule.

It is noted that in the thermodynamic limit, the binding reaction is described by the Michaelis--Menten equation for the timescale separation.
The Michaelis--Menten equation for multiple substrates with the conservation of the enzyme concentration is given as
\begin{equation}
N^{\mathrm{tot}}_E =\frac{N_{S0} N^{\mathrm{free}}_E}{K_0 + N^{\mathrm{free}}_E} + \frac{N_{S1} N^{\mathrm{free}}_E}{K_1 + N^{\mathrm{free}}_E} + N^{\mathrm{free}}_E, \label{eq:MM}
\end{equation}
where $N^{\mathrm{tot}}_E$ is the concentration of all enzymes, $N^{\mathrm{free}}_E$ is the concentration of free enzymes that do not bind to any substrate, $N_{Si}$ is the concentration of the $i$-th substrate, and $K_i$ is the dissociation constant between the enzyme and the $i$-th substrate.
A comparison of Eqs.~(\ref{eq:avg_n}) and (\ref{eq:MM}) indicates that $\exp(-\beta \epsilon^{\rm b}_{\sigma})$ is $K_\sigma$, $\exp(\beta \mu)$ is $N^{\mathrm{free}}_E$, and $<n>$ is $N^{\mathrm{tot}}_E - N^{\mathrm{free}}_E$.
If $N^{\mathrm{tot}}_E$ is smaller than $N_{S0} + N_{S1}$ and $K_i$ is much smaller than $S_i$, almost all enzymes bind to substrates; that is, $<n> \simeq N^{tot}_E$.
Under this condition, $N_{S0} N^{\mathrm{free}}_E / (K_0 + N^{\mathrm{free}}_E)$ and $N_{S1} N^{\mathrm{free}}_E / (K_1 + N^{\mathrm{free}}_E)$ have to be $\mathcal{O}(N^{\mathrm{tot}}_E)$, where $\mathcal{O}$ is the Landau notation.
Thus, $N^{\mathrm{free}}_E$ is estimated as $N^{\mathrm{free}}_E / N^{\mathrm{tot}}_E \sim K_i / N_{Si}$.
Thus, when the substrates have sufficiently large binding energies for the enzyme, $N^{\mathrm{free}}_E$ is negligible compared with $N^{\mathrm{tot}}_E$, and $<n>$ is considered to be constant throughout the relaxation process for actual {\it in vivo} and {\it in vitro} reactions.
In contrast, when $N^{\mathrm{tot}}_E$ is higher than $N_{S0} + N_{S1}$, $<n>/N$ is 1 through the relaxation process for a sufficiently small binding energy.

\section{Variance among samples}

Interestingly, the variance of $<\mathcal{U}>_{\rm ens}$ or $<\mathcal{T}>_{\rm ens}$ with the samples increased at the initiation of the plateaus (Fig. S\ref{fig:spin_var_time_course}).
The maximum variance, which was normalized by division by the average over the  samples, increased gradually with decreasing temperature (see Fig. S\ref{fig:spin_var_time_course}(a)). This corresponded to the gradual appearance of the plateau in the averaged relaxation dynamics (Fig. 2(a) in the main text).
As $<n>$ decreased, the change in the variance became more drastic.
When $<n>/N$ was 1, there was only one peak at an early step (see the purple line in Fig. S\ref{fig:spin_var_time_course}(b)).
At $<n>/N = 0.9$, a second plateau appeared suddenly in the average relaxation dynamics (see Fig. 2(b) in the main text). Accordingly, a second peak in the variance appeared suddenly (see the blue line in Fig. S\ref{fig:spin_var_time_course}(b)).
The variance increased at the beginning of the plateaus because the number of modified spins varied with each sample, and the induced variance froze at the plateaus (see Fig. S\ref{fig:var_mol_num}).
Therefore, the variance among samples provided a good indicator of the plateaus.

To clarify the region where multiple plateaus appear, we plotted the maximum variance after the first peak by subtracting the equilibrium value of the variance (Fig. S\ref{fig:spin_var_diagram}(b)).
The diagram shows that the variance was high under low-$<n>$ and low-temperature conditions, which indicates the existence of multiple plateaus similar to glassy dynamics.

\section{Supporting figures}

\begin{figure}[htbp]
\begin{center}
\includegraphics{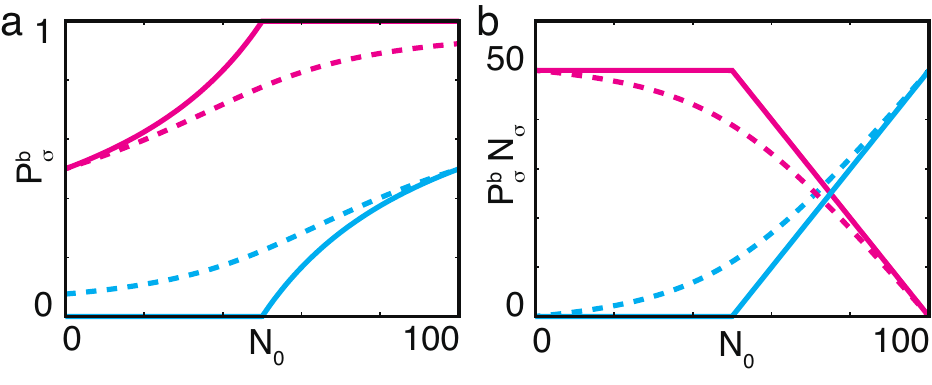}
\caption{
Binding probability between the enzyme and the substrates.
(a) Binding probability ($P^{\rm b}_{\sigma}$) and (b) the number of complexes ($P^{\rm b}_{\sigma} N_{\sigma}$) for different substrates at different temperatures.
Cyan and magenta represent $\sigma = 0$ and $\sigma = 1$ substrates, respectively.
Solid and dashed lines represent $\beta =$ 1.75 and 0.25, respectively.
$N$ and $<n>/N$ were set to 100 and 0.5, respectively.
\label{fig:bind_prob}
}
\end{center}
\end{figure}

\begin{figure}[htbp]
\begin{center}
\includegraphics[width=6.0cm]{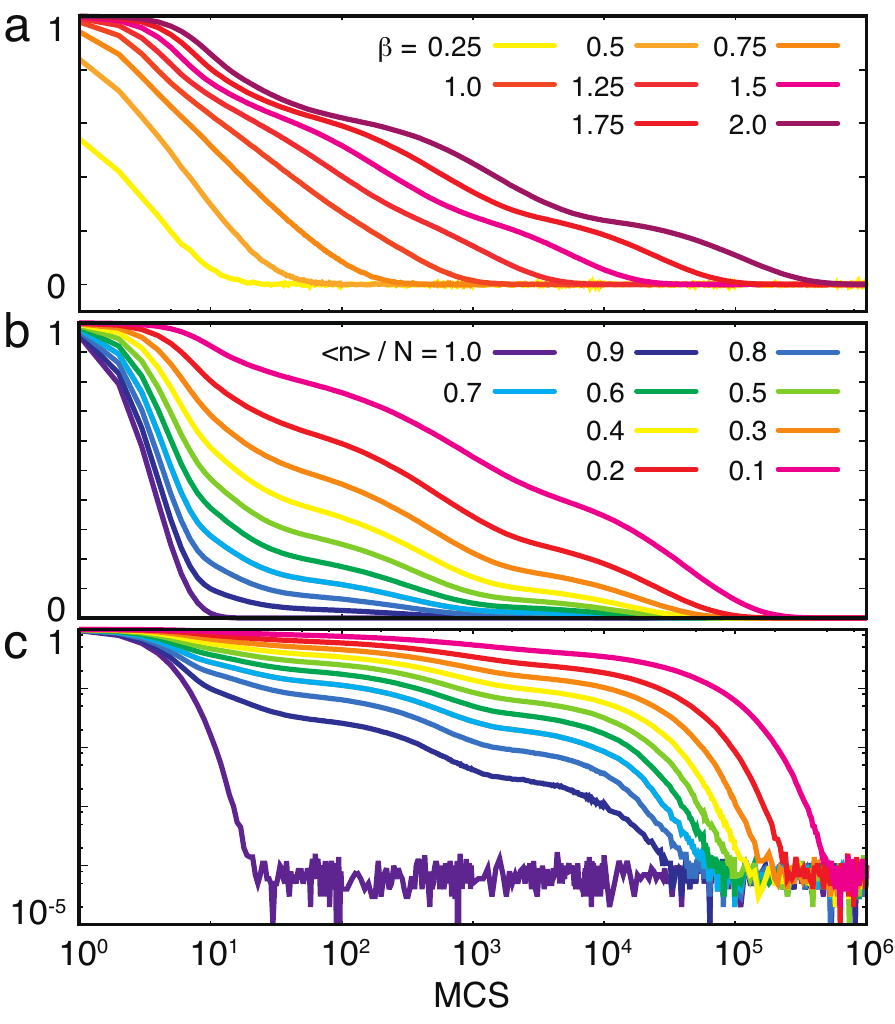}
\caption{
Time evolutions of the average structural spin in the enzymatic MWC model against logarithmic time (Monte Carlo step).
(a) Relaxation of the average number of structural spins to the equilibrium state at various temperatures.
The equilibrium fraction of T-state molecules $\mathcal{T}_{\rm eq}$ depends on the temperature; therefore, the normalized ratio $(<\mathcal{T}>_{\rm ens} - \mathcal{T}_{\rm eq})/(1 - \mathcal{T}_{\rm eq})$ was plotted.
$<n>/N$ was set to 0.2.
The different line colors show the time courses at different $\beta$ values.
(b) Relaxation of the average spin state to the equilibrium state at various values of $<n>$.
The equilibrium fraction of T-state molecules did not change with the changes in $<n>$, and $<\mathcal{T}>_{\rm ens}$ was plotted.
$\beta$ was set to 1.75.
The different line colors indicate the time courses with different $<n>$ values.
Each line is an ensemble average of 1000 samples.
(c) Logarithm of $<\mathcal{T}>_{\rm ens}$ plotted for (b).
\label{fig:state_time_course}
}
\end{center}
\end{figure}

\begin{figure}[htbp]
\begin{center}
\includegraphics[width=6.0cm]{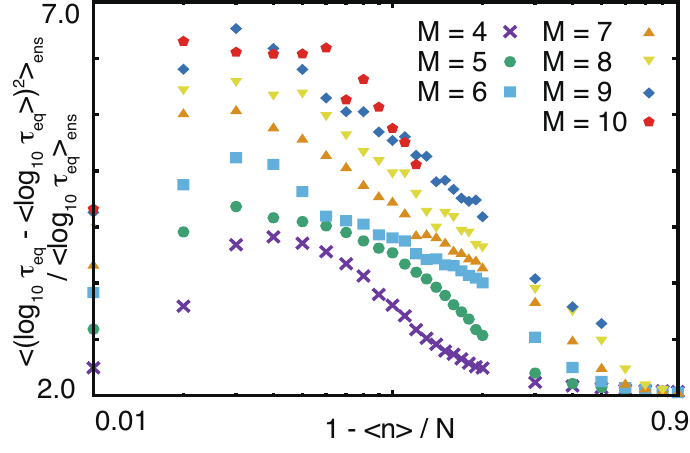}
\caption{
The variance of logarithmic relaxation time for various values of $M$.
The relaxation time of each sample was calculated in the same manner as in Fig.~4 in the main text.
The variance was normalized by the averaged value.
The horizontal axis denotes the difference of $<n>/N$ from 1.0 and the vertical axis denotes the variance of the logarithmic relaxation time.
Each symbol indicates the variance for a different value of the number of modification sites $M$. 
}
\end{center}
\end{figure}

\begin{figure}[htbp]
\begin{center}
\includegraphics[width=12.0cm]{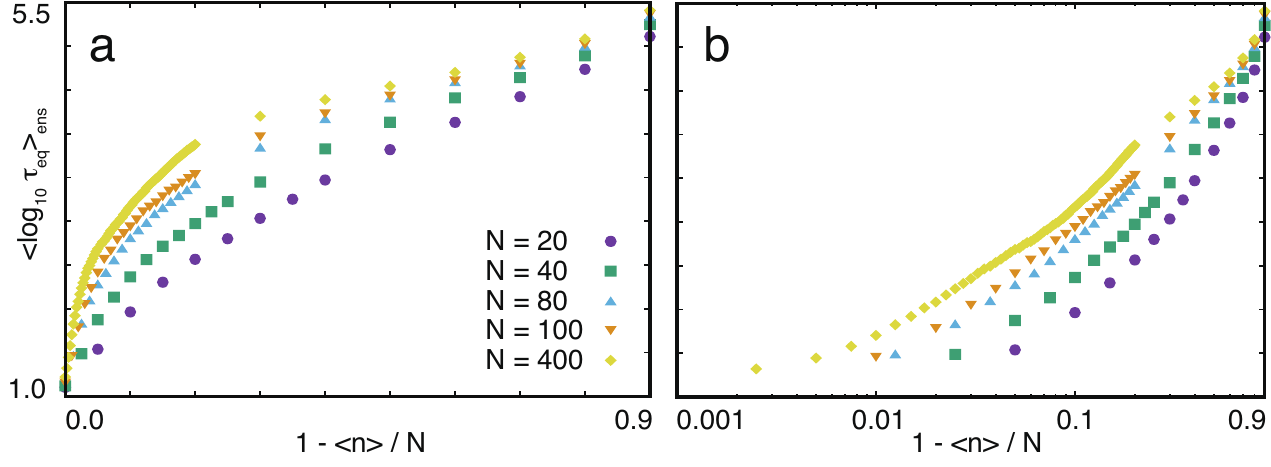}
\caption{
The average of logarithmic relaxation time plotted against $N$.
The relaxation time of each sample was calculated in the same manner as in Fig.~4 in the main text.
The horizontal axis denotes the difference of $<n>/N$ from 1.0 (a) and its logarithmic plot (b).
Different symbols indicate relaxation time for systems with different number of molecules $N$.
}
\end{center}
\end{figure}

\begin{figure}[htbp]
\begin{center}
\includegraphics[width=12.0cm]{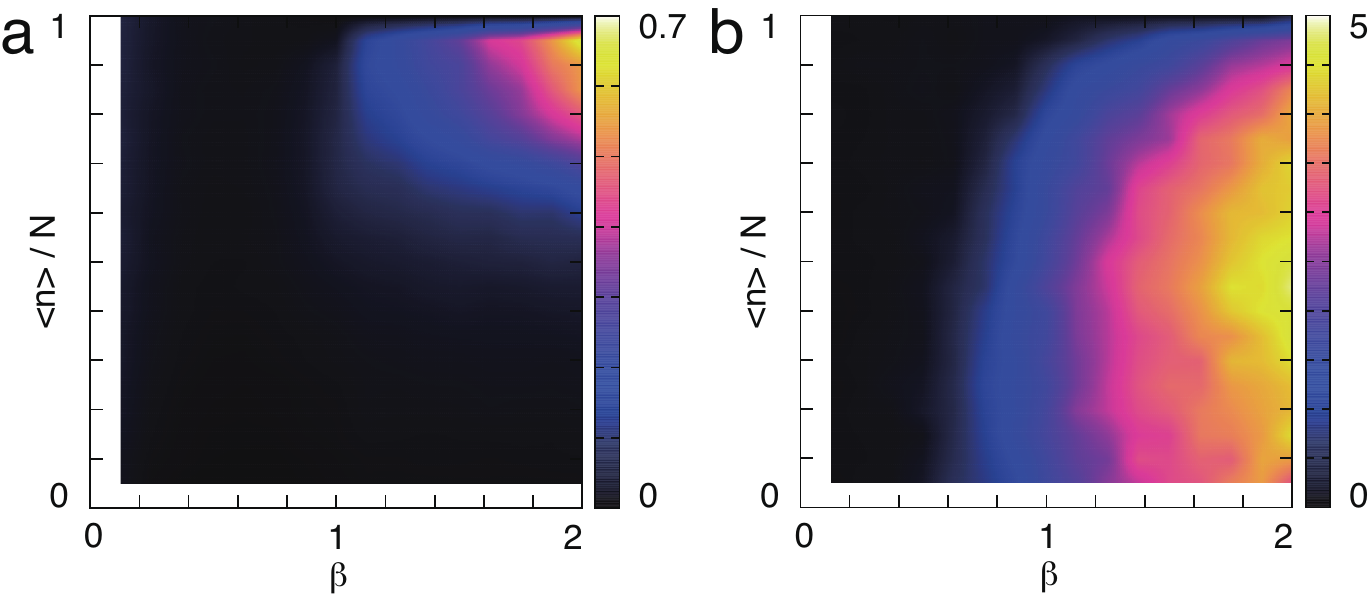}
\caption{
Color maps of the variance of the logarithmic relaxation time and the number of unmodified monomers.
(a) $<[\log_{10} \tau_{\rm eq} - <\log_{10} \tau_{\rm eq}>]^2>_{\rm ens}$ against $\beta$ and $<n>/N$.
The variance is indicated by the colors in the side bar.
\label{fig:relax_time_dist}
(b) Maximal variance in the number of unmodified monomers.
This was calculated as the maximum variance among the ensemble in the time course after 20 MCSs.
The logarithmic relaxation time was divided by the average $<\mathcal{U}>_{\rm ens}$ and the result was subtracted from the analytically calculated value $<(\mathcal{U} - \mathcal{U}_{\rm eq})^2>_{\rm eq}/\mathcal{U}_{\rm eq}$.
\label{fig:spin_var_diagram}
}
\end{center}
\end{figure}

\begin{figure}[htbp]
\begin{center}
\includegraphics[width=6.0cm]{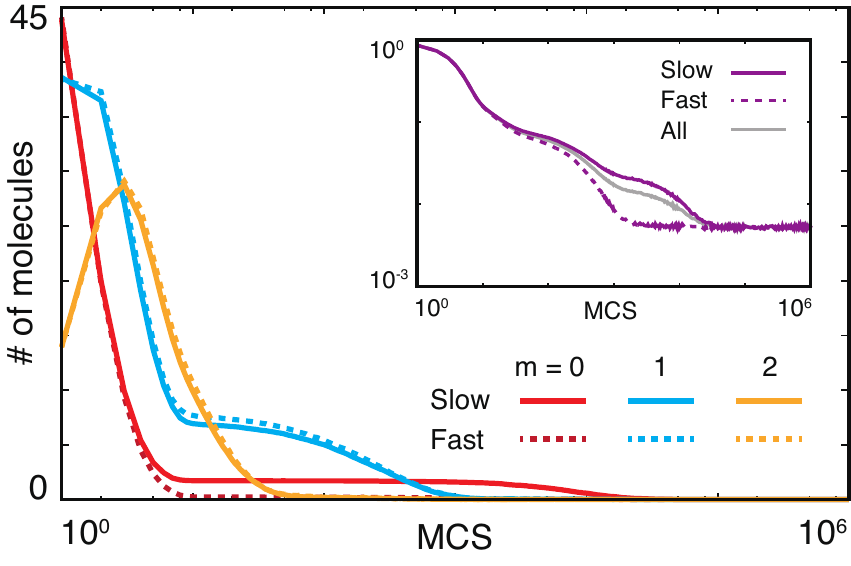}
\caption{
Averaged dynamics for fast and slow groups in the bimodal distribution of the relaxation time with $<n>/N = 0.8$ and $\beta = 1.75$.
The relaxation time of the samples in the fast group was faster than $10^{3.3}$ MCSs, and that in the slow group was slower than $10^{3.3}$ MCSs.
The fast and slow groups had 416 and 584 samples, respectively.
The red, blue, and yellow lines indicate the dynamics of molecules with zero, one, and two modified monomers, respectively.
The dotted and solid lines indicate the dynamics of the fast and slow groups, respectively.
Inset: Relaxation of the number of unmodified monomers $\mathcal{U}$ of the fast and slow groups.
Purple lines indicate the dynamics of each group, and the grey line indicates the dynamics of the whole sample.
\label{fig:fast_slow}
}
\end{center}
\end{figure}

\begin{figure}[htbp]
\begin{center}
\includegraphics[width=6.0cm]{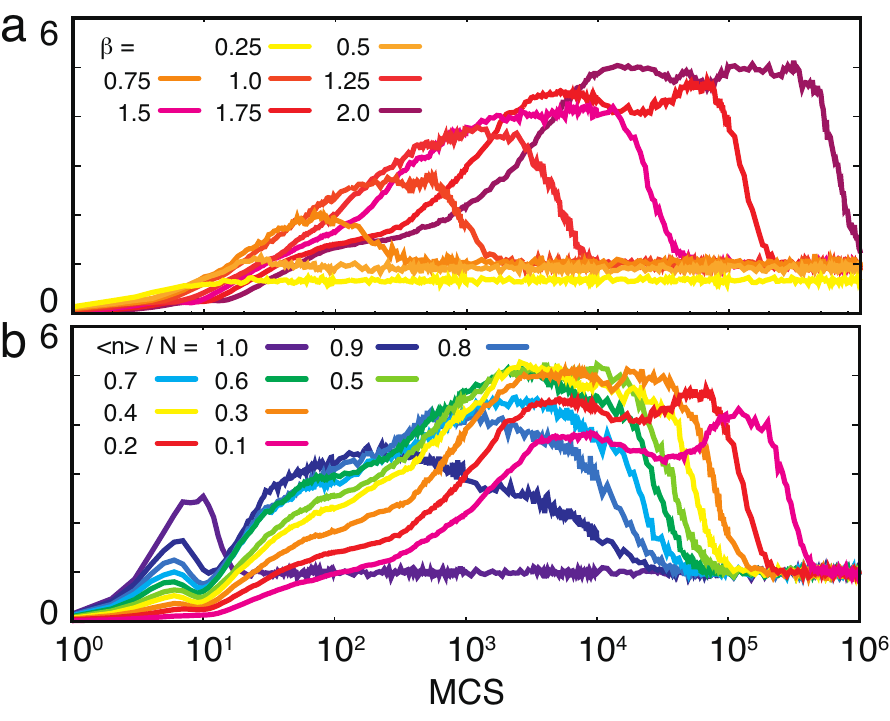}
\caption{
Time evolution of the variance in the number of modification spins toward equilibrium in the enzymatic MWC model.
(a) Relaxation of the variance in the number of modification spins to the equilibrium state at various temperatures.
The variance was normalized by dividing it by the average.
$<n>/N$ was set at 0.2.
The different line colors indicate the time courses with different $\beta$ values.
(b) Relaxation of the variance in the state spin to the equilibrium state at various $<n>$ values.
$\beta$ was set as 1.75.
The different line colors indicate the time courses with different $<n>/N$ values.
\label{fig:spin_var_time_course}
}
\end{center}
\end{figure}

\begin{figure}[htbp]
\begin{center}
\includegraphics[width=6.0cm]{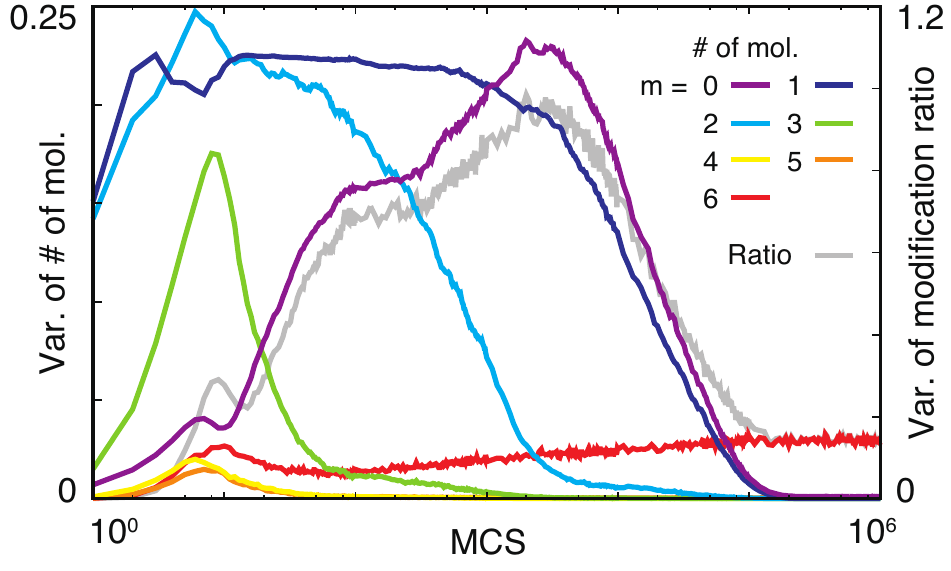}
\caption{
Time evolution of the variance in the number of molecules with different modification spins at $<n>/N = 0.2$ and $\beta = 1.75$.
The grey line indicates the variance in $\mathcal{U}$, and the other colored lines show the variance in the number of molecules normalized by division by $N$.
The different colors indicate molecules with different numbers of modified monomers.
Purple, blue, cyan, green, yellow, orange, and red indicate molecules with $m =$ 0, 1, 2, 3, 4, 5, and 6, respectively.
\label{fig:var_mol_num}
}
\end{center}
\end{figure}